\documentclass[sigconf]{acmart}
\AtBeginDocument{%
  \providecommand\BibTeX{{%
    \normalfont B\kern-0.5em{\scshape i\kern-0.25em b}\kern-0.8em\TeX}}}
\usepackage{xspace}
\usepackage{amsthm}
\usepackage{amsmath}
\usepackage{enumitem}
\usepackage{booktabs}
\usepackage{multirow}
\usepackage{graphicx}
\usepackage[normalem]{ulem}
\useunder{\uline}{\ul}{}
\usepackage{caption}
\usepackage{subcaption}
\usepackage{balance}
\newcommand{\modelname}{\textsf{DT4SR}\xspace}
\setcopyright{acmcopyright}
\copyrightyear{2018}
\acmYear{2018}
\acmDOI{10.1145/1122445.1122456}

\acmConference[Woodstock '18]{Woodstock '18: ACM Symposium on Neural
  Gaze Detection}{June 03--05, 2018}{Woodstock, NY}
\acmBooktitle{Woodstock '18: ACM Symposium on Neural Gaze Detection,
  June 03--05, 2018, Woodstock, NY}
\acmPrice{15.00}
\acmISBN{978-1-4503-XXXX-X/18/06}

\begin{document}

%%
%% The "title" command has an optional parameter,
%% allowing the author to define a "short title" to be used in page headers.
\title{Modeling Sequences as Distributions with Uncertainty for Sequential Recommendation}

%%
%% The "author" command and its associated commands are used to define
%% the authors and their affiliations.
%% Of note is the shared affiliation of the first two authors, and the
%% "authornote" and "authornotemark" commands
%% used to denote shared contribution to the research.
\author{Ziwei Fan, Zhiwei Liu}
% \thanks{*Both authors contribute equally.}
% \email{zliu213@uic.edu}
\affiliation{%
  \institution{Department of Computer Science, University of Illinois at Chicago}
  \country{USA}
}
\email{{zliu213,zfan20}@uic.edu}
% \author{Ziwei Fan}
% % \authornote{Both authors contributed equally to this research.}
% \email{trovato@corporation.com}
% \orcid{1234-5678-9012}
% \author{G.K.M. Tobin}
% \authornotemark[1]
% \email{webmaster@marysville-ohio.com}
% \affiliation{%
%   \institution{Institute for Clarity in Documentation}
%   \streetaddress{P.O. Box 1212}
%   \city{Dublin}
%   \state{Ohio}
%   \country{USA}
%   \postcode{43017-6221}
% }

\author{Lei Zheng}
\affiliation{%
  \institution{Pinterest Inc.}
%   \streetaddress{1 Th{\o}rv{\"a}ld Circle}
%   \city{Hekla}
%   \country{Iceland}}
    \country{USA}
}
\email{lzheng@pinterest.com}

\author{Shen Wang, Philip S. Yu}
\affiliation{%
  \institution{Department of Computer Science, University of Illinois at Chicago}
  \country{USA}
%   \streetaddress{1 Th{\o}rv{\"a}ld Circle}
%   \city{Hekla}
%   \country{Iceland}}
}
\email{{swang224,psyu}@uic.edu}

% \author{Valerie B\'eranger}
% \affiliation{%
%   \institution{Inria Paris-Rocquencourt}
%   \city{Rocquencourt}
%   \country{France}
% }

% \author{Aparna Patel}
% \affiliation{%
%  \institution{Rajiv Gandhi University}
%  \streetaddress{Rono-Hills}
%  \city{Doimukh}
%  \state{Arunachal Pradesh}
%  \country{India}}

% \author{Huifen Chan}
% \affiliation{%
%   \institution{Tsinghua University}
%   \streetaddress{30 Shuangqing Rd}
%   \city{Haidian Qu}
%   \state{Beijing Shi}
%   \country{China}}

% \author{Charles Palmer}
% \affiliation{%
%   \institution{Palmer Research Laboratories}
%   \streetaddress{8600 Datapoint Drive}
%   \city{San Antonio}
%   \state{Texas}
%   \country{USA}
%   \postcode{78229}}
% \email{cpalmer@prl.com}

% \author{John Smith}
% \affiliation{%
%   \institution{The Th{\o}rv{\"a}ld Group}
%   \streetaddress{1 Th{\o}rv{\"a}ld Circle}
%   \city{Hekla}
%   \country{Iceland}}
% \email{jsmith@affiliation.org}

% \author{Julius P. Kumquat}
% \affiliation{%
%   \institution{The Kumquat Consortium}
%   \city{New York}
%   \country{USA}}
% \email{jpkumquat@consortium.net}

%%
%% By default, the full list of authors will be used in the page
%% headers. Often, this list is too long, and will overlap
%% other information printed in the page headers. This command allows
%% the author to define a more concise list
%% of authors' names for this purpose.
\renewcommand{\shortauthors}{Trovato and Tobin, et al.}

%%
%% The abstract is a short summary of the work to be presented in the
%% article.
\begin{abstract}
  The sequential patterns within the user interactions are pivotal for representing the user's preference and capturing latent relationships among items. The recent advancements of sequence modeling by Transformers advocate the community to devise more effective encoders for the sequential recommendation. Most existing sequential methods assume users are deterministic. However, item-item transitions might fluctuate significantly in several item aspects and exhibit randomness of user interests. This \textit{stochastic characteristics} brings up a solid demand to include uncertainties in representing sequences and items. Additionally, modeling sequences and items with uncertainties expands users' and items' interaction spaces, thus further alleviating cold-start problems.
%   We argue that users with fluctuating interests are more uncertain and challenging to model dynamic preferences. This brings up a solid demand to include uncertainties in representations of sequences and items. We also suggest that modeling sequences and items with uncertainty is beneficial for cold-start user and item problems.

%   Most existing sequential methods represent each user\slash item with one point embedding vector and use dot-product to express the affinity between users and items. However, these methods fail to fully recognize and utilize two characteristics in sequential recommendation: (1). a recommendation model satisfying triangle inequality provides better generalization and reasoning ability when data is limited~(cold start problem); (2). users with diverse interests are typically more uncertain than single-minded users, thus, representations with uncertainty are necessary.
  
  In this work, we propose a \textbf{D}istribution-based \textbf{T}ransformer for \textbf{S}equentail \textbf{R}ecommendation (\modelname), which 
%   to 
%   alleviate aforementioned issues in sequential recommendation,
    injects uncertainties into sequential modeling.
%   for better representations of
% to learn
%   items and sequences embeddings.
  We use Elliptical Gaussian distributions to describe items and sequences with uncertainty.
  We describe the uncertainty in items and sequences as Elliptical Gaussian distribution. 
  And we adopt Wasserstein distance to measure 
%   how likely each item-item transition is
the similarity between distributions. We devise two novel Transformers for modeling mean and covariance, which
%   are proposed so that the
guarantees the positive-definite 
  property of distributions.
%   are guaranteed.
  The proposed method significantly outperforms the state-of-the-art methods. The experiments on three benchmark datasets also demonstrate its effectiveness in alleviating cold-start issues. The code is available in \url{https://github.com/DyGRec/DT4SR}.
%   cold-start user\slash item problems. 
\end{abstract}

%%
%% The code below is generated by the tool at http://dl.acm.org/ccs.cfm.
%% Please copy and paste the code instead of the example below.
%%
% \begin{CCSXML}
% <ccs2012>
%  <concept>
%   <concept_id>10010520.10010553.10010562</concept_id>
%   <concept_desc>Computer systems organization~Embedded systems</concept_desc>
%   <concept_significance>500</concept_significance>
%  </concept>
%  <concept>
%   <concept_id>10010520.10010575.10010755</concept_id>
%   <concept_desc>Computer systems organization~Redundancy</concept_desc>
%   <concept_significance>300</concept_significance>
%  </concept>
%  <concept>
%   <concept_id>10010520.10010553.10010554</concept_id>
%   <concept_desc>Computer systems organization~Robotics</concept_desc>
%   <concept_significance>100</concept_significance>
%  </concept>
%  <concept>
%   <concept_id>10003033.10003083.10003095</concept_id>
%   <concept_desc>Networks~Network reliability</concept_desc>
%   <concept_significance>100</concept_significance>
%  </concept>
% </ccs2012>
% \end{CCSXML}

% \ccsdesc[500]{Computer systems organization~Embedded systems}
% \ccsdesc[300]{Computer systems organization~Redundancy}
% \ccsdesc{Computer systems organization~Robotics}
% \ccsdesc[100]{Networks~Network reliability}

%%
%% Keywords. The author(s) should pick words that accurately describe
%% the work being presented. Separate the keywords with commas.

%% A "teaser" image appears between the author and affiliation
%% information and the body of the document, and typically spans the
%% page.

%%
%% This command processes the author and affiliation and title
%% information and builds the first part of the formatted document.
\maketitle

\section{Introduction}

\begin{figure}
    \centering
    \includegraphics[width=0.45\textwidth]{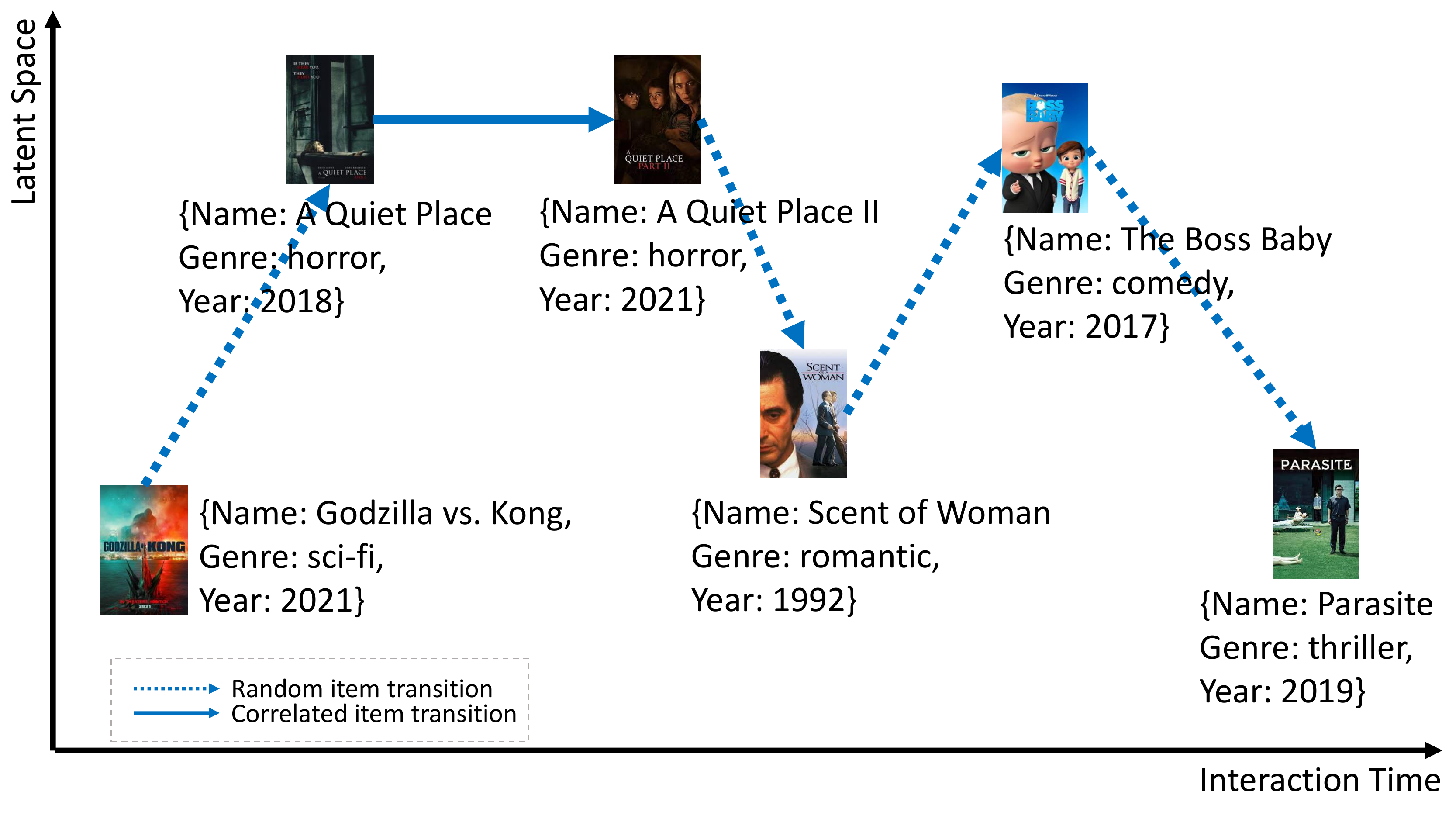}
    % \vspace{-5mm}
    \caption{An example user with uncertain interests. The random item transition connects two items with distinct genres and release years, while a correlated item transition connects items with same genres and/or release years. }
    \label{fig:motivation}
\end{figure}
% \vspace{-10mm}
% Selling points:
% \begin{itemize}
%     \item Triangular inequality for data sparsity and better item representations learning in sequential learning.
%     \item Uncertainty within item representations.
% \end{itemize}

Recommendation systems achieve great successes in providing personalized services in various domains, including fashion~\cite{kang2017visually}, music~\cite{van2013deep}, book~\cite{givon2009predicting} and grocery~\cite{faggioli2020recency,liu2020basket,liu2020basconv} recommendation. 
% The fundamental idea of recommendation is Collaborative Filtering~(CF)~\cite{wang2019neural,rendle2012bpr,koren2009matrix,he2020lightgcn,tay2018latent} suggesting that similar users have similar tastes. 
% CF approaches aim at understanding users' preferences and latent concepts of items by modeling user-item interactions. 
Among recommendation methods, Sequential Recommendation~(SR) methods~\cite{tang2018personalized, chen2018sequential, kang2018self, sun2019bert4rec, zheng2019gated} show promising improvements in predicting users' interests. 
SR methods format each user's historically interacted items as a sequence and 
dynamically
model the user's preference 
% dynamics
% for next-item prediction.
to predict the next item in sequences.
% , especially Transformer~\cite{vaswani2017attention}-based methods~\cite{kang2018self, sun2019bert4rec}. 
% Typical recommendation methods learn the user-item affinity based on known interactions, while
% SR methods format each user's historically interacted items as a sequence and model the user's preference dynamics for next-item prediction. 

The core idea of SR is modeling item-item transition
relationships 
within the sequence. 
The recent advancements of Transformer~\cite{vaswani2017attention} on sequence encoding inspire the recommendation community to adopt it for user sequences modeling~\cite{kang2018self, sun2019bert4rec}.
SASRec~\cite{kang2018self} is 
% the
a pioneering work proposing to use Transformer for recommendation
% . It
, which
applies self-attention 
% via scaled dot-product 
to measure the correlation among item transitions. 
Several following 
% work
works
~\cite{sun2019bert4rec, li2020time, ma2019hierarchical} 
% proposes
% several improvements
enhance the SASRec model with complex components to improve the recommendation performance,
% building upon~\cite{kang2018self} and
which 
demonstrates the effectiveness of
adopting
Transformer as a backbone encoder for sequence modeling.

% However, the interests of users fluctuate and sometimes hard to explain. In other words, they are not deterministic~\cite{anderson2020algorithmic}. 
However, the \textit{stochastic characteristics} of item transitions in sequences spoil the ability of existing models to model sequential correlations,
which is demonstrated in 
% For example, 
Figure~\ref{fig:motivation}. In this Figure, we present the movie-watching records of a user
and the corresponding genre and release year transitions 
between items.
% presents a user's interaction sequence.
We can observe that 
% most 
item-item transition 
relationships
% of this user 
fluctuate 
% significantly, 
randomly
% in
from
both genre and release year perspectives.
% ranging from a sci-fi movie ``Godzilla vs. Kong'', horror movies ``A Quiet Place'', and a romantic old movie ``Scent of a Woman''. 
Most item-item transitions are hard to explain as both items in a transition pair
have distinct genres and years, 
% do not belong to the same category or are not even produced in the same year,
% such as
\textit{e.g.,}
the transition from ``\textit{Scent of a Woman}'' to ``\textit{The Boss Baby}''. 
% However, some of them are still easily concluded, such as items of second item-item transition ``A Quiet Place I'' to ``A Quiet Place II'' are in the same genre and same series.
This example 
% verifies 
indicates
the importance of modeling each sequence with \textbf{uncertainty}. For each user, we 
% can 
should
consider her interactions as a set of stochastic events controlled by distributions of dynamic user interests.
% As to sequential recommendation, 
Modeling sequence as distributions can not only incorporate such transition uncertainty but also benefit the exploration of users' interests. 
The reason is that a distribution covers more data space than an embedding, which thus expands the interaction space of a user. As such, the user cold-start issue, \textit{i.e.,} users with few interactions~\cite{liu2021augmenting}, is alleviated.    

Previous works~\cite{oh2018modeling,zheng2019deep} have demonstrated the superiority of representing a stochastic object as a distribution rather than an embedding.
% embeddings of uncertainty.
% Modeling embeddings with uncertainty~\cite{oh2018modeling} 
% demonstrates its superiority in representing ambiguous objects.
% In SR, when users' interests significantly change in and the underlying intentions of item-item transitions are unknown, uncertainty becomes crucial.
% However, it is challenging to represent items and sequences~(i.e., users) with uncertainty. In the meantime, it is unclear how to measure the similarity between items with uncertainty.
Nevertheless, representing items and their associated transitions 
% with 
as
distributions is challenging. Firstly, it is non-trivial to dynamically learn mean and covariance to infer the distribution. 
% are 
% learned in a dynamic sequence modeling setting. 
Moreover, we need to guarantee the positive definite property of covariance during inference. Furthermore, it is still unclear how to measure the distance between the sequence distributions.
% the calculation of latent similarity of items in the same transition remains unclear. 

% To this end, we use probabilistic distributions to represent items and further infer sequences as distributions and describe user's interests. 
To this end, we propose a distribution-based Transformer~(\modelname) to model the dynamics of evolving distribution representations while still maintaining necessary properties of distribution.
To be specific, instead of a fixed vector, we use Elliptical Gaussian distributions~\cite{vilnis2014word,bojchevski2017deep, zheng2019deep} to represent items with covariance measuring the uncertainty. We develop mean and covariance Transformers to learn the dynamics of mean and covariance in the user sequence. Instead of using dot-product to measure the affinity between the inferred next-item and item candidates, we propose to use Wasserstein distance under a metric learning framework~\cite{hsieh2017collaborative, li2020symmetric, tay2018latent} to generate a top-N recommendation list of items with the smallest distribution distances.
The followings are our contributions:
\begin{itemize}[leftmargin=*]
    \item To the best of our knowledge, this is the first work proposing to model items and sequences as Elliptical Gaussian Distributions in the sequential recommendation. We demonstrate that distribution representations with uncertainty well characterize fluctuating and evolving interests of users and further alleviate cold-start problem.
    % We demonstrate that distribution-based sequential recommendation helps both overall performance and cold start users\slash items.
    \item We develop two novel Transformers for modeling mean and covariance embeddings adaptive to the distribution-based sequential recommendation.
    \item The \modelname achieves significant improvements over state-of-the-art recommendation methods in both overall performance and cold start setting. The experimental results verify the effectiveness of distribution-based representations in sequential modeling.
\end{itemize}

\section{Problem Definition}
A 
% typical 
recommender system collects feedback 
% from
between
a set of users $\mathcal{U}$ and items $\mathcal{V}$, 
% in various forms, 
% such as ratings and
\textit{e.g.,}
clicks. 
% We denote each user as $u$ and each item as $v$. 
Sequential recommendation chronologically models a user $u$'s interaction sequence,
% takes each user's historical interactions as the input and sort them chronologically to form a sequence,
% where we use
which is denoted as 
$\mathcal{S}^{u}={[v^{u}_1,v^u_2,\dots,v^u_{|\mathcal{S}^{u}|}]}$. 
% to denote user $u$'s sequence.
The goal of SR is to 
% model and capture 
characterize
dynamics in sequences
% from her sequence 
and then predict the 
% potential 
next-item.
% to interact with
% in the sequence 
We formulate the objective as follows:
\begin{equation}
    p\left( v_{|\mathcal{S}^{u}|+1}^{(u)}=v \left|  \mathcal{S}^{u} \right.\right),
\end{equation}
which 
% measuring 
measures
the probability of an item $v$ being the next item, given user $u$'s sequence $\mathcal{S}^{u}$.

\section{Proposed Model}
\begin{figure}
    \centering
    \includegraphics[width=0.5\textwidth]{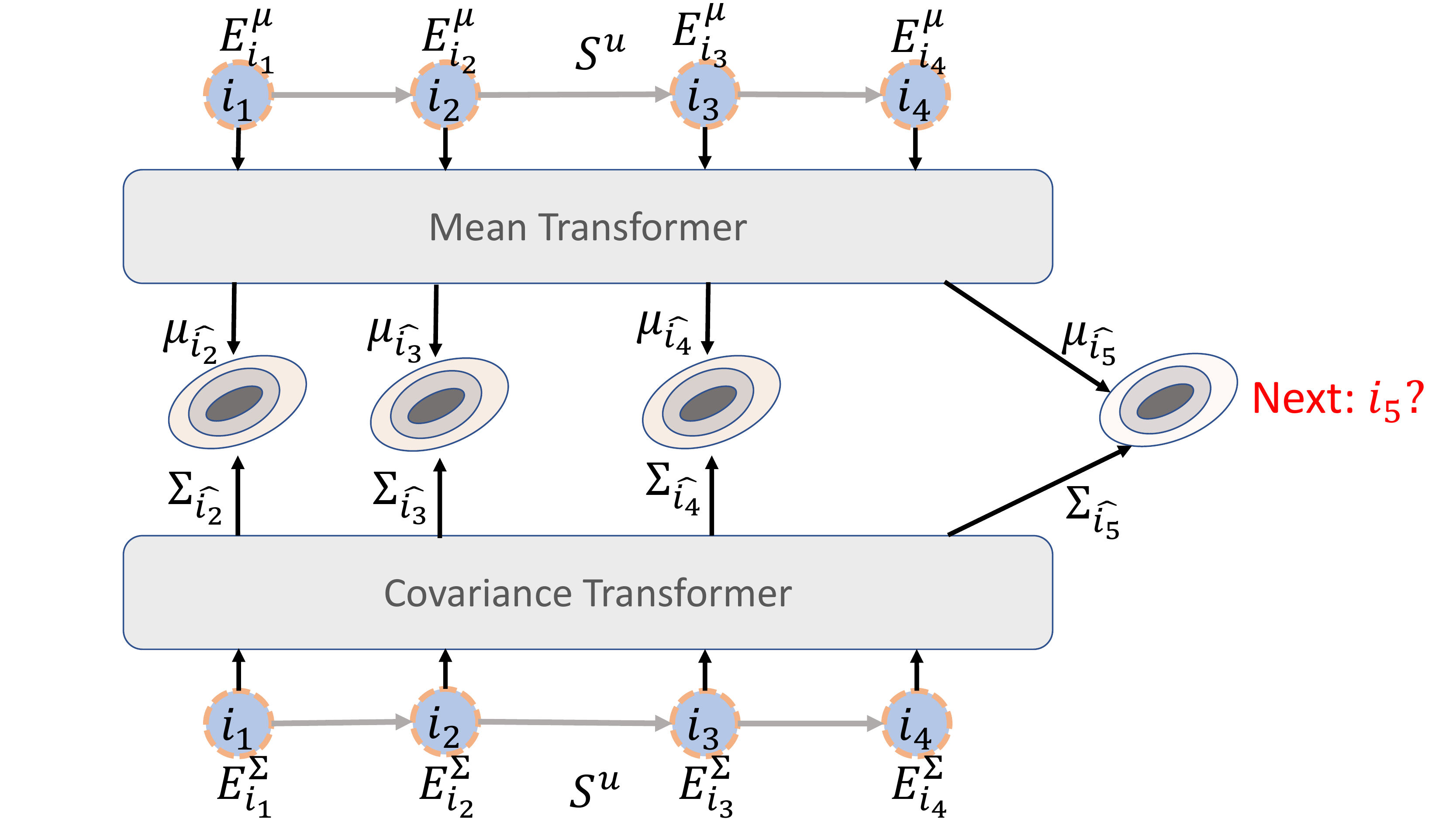}
    % \vspace{-5mm}
    \caption{: Illustration of the proposed \modelname. Mean and Covariance Transformers take the sequences of mean and covariance embeddings as inputs and output next-step item's distribution. For example, at step $1$, $\mathbf{E}^{\mu}_{i_1}$ and $\mathbf{E}^{\Sigma}_{i_1}$ together infer $\mathbf{E}^{\mu}_{\hat{i_2}}$ and $\mathbf{E}^{\Sigma}_{\hat{i_2}}$.}
    \label{fig:framework}
\end{figure}

In this section, we present the framework of \modelname, as shown in Figure~\ref{fig:framework}. The overall framework consists of several components, \textit{mean and covariance embeddings}, \textit{distribution-based self-attention}, \textit{distribution-based Feed-Forward Network}~(FFN), and a specific \textit{covariance output layer}. The \textit{mean and covariance embeddings} together describe distributions of items. The \textit{distribution-based self-attention} 
% attentively 
captures dynamical correlations in mean and covariance aspects. The \textit{distribution-based FFN} introduces non-linearity specifically for distribution representation. To guarantee the positive definite property of covariance, we propose a \textit{covariance output layer}.
% The mean and covariance embeddings are randomly initialized and learned after optimization. The output mean and covariance of both Transformers at each step of sequence together describe the distribution of next-step interacted item. 
% The optimized Transformers will generate final step's mean and covariance vectors for predicting next preferred item's distribution. 
This distribution-based dynamic modeling distinguishes our proposed method from existing sequential methods.

\subsection{Embedding Layers}
% Different from existing sequential methods, we use two embedding tables for all items. 
We represent each item with an elliptical Gaussian distribution governed by a mean vector and one diagonal covariance matrix. To be specific, each item has two embedding representations, which are for mean and covariance.
We denote the mean embedding table as $\mathbf{E}^{\mu}\in \mathbb{R}^{|\mathcal{V}|\times d}$ and covariance embedding table as $\mathbf{E}^{\Sigma}\in \mathbb{R}^{|\mathcal{V}|\times d}$, where $\mathcal{V}$ denotes the item set and $d$ is the embedding dimensionality. We also define separate learnable positional embeddings for both Transformers, $\mathbf{P}^{\mu}\in \mathbb{R}^{n\times d}$ for the mean Transformer and $\mathbf{P}^{\Sigma}\in \mathbb{R}^{n\times d}$ for the covariance Transformer respectively. The $n$ denotes the maximum length of the sequence. The elements $p_i$ signifies the positional information at position $i$ in the sequence. As the sequence lengths of users vary a lot, we keep the most recent $n$ interactions to format the sequence if a user has more than $n$ interactions. Otherwise, we keep all interactions and apply zero paddings until the sequence has a length of $n$. With all embeddings, given a sequence $\mathcal{S}={[v_1,v_2,\dots,v_n]}$, we have its mean and covariance sequence representations as:
\begin{equation}
\begin{aligned}
    \label{eq:sequence_embed}
    \mathbf{E}^{\mu}_{\mathcal{S}} &= [\mathbf{e}^{\mu}_1+\mathbf{p}^{\mu}_1, \mathbf{e}^{\mu}_2+\mathbf{p}^{\mu}_2, \dots, \mathbf{e}^{\mu}_n+\mathbf{p}^{\mu}_n],\\
    \mathbf{E}^{\Sigma}_{\mathcal{S}} &= [\mathbf{e}^{\Sigma}_1+\mathbf{p}^{\Sigma}_1, \mathbf{e}^{\Sigma}_2+\mathbf{p}^{\Sigma}_2, \dots, \mathbf{e}^{\Sigma}_n+\mathbf{p}^{\Sigma}_n].
\end{aligned}
\end{equation}

\subsection{Mean and Covariance Self-Attentions}
We introduce a novel distribution-based self-attention mechanism considering numerical stability and its applicability to distribution embeddings.
A typical self-attention block introduces query~$\mathbf{Q}$, key~$\mathbf{K}$, and value~$\mathbf{V}$. In sequential recommendation, the query~$\mathbf{Q}$, key~$\mathbf{K}$, and value~$\mathbf{V}$ are linear transformations of the same sequence embedding~$\mathbf{E}_{\mathcal{S}}^{*}\mathbf{W}^{Q}$, $\mathbf{E}_{\mathcal{S}}^{*}\mathbf{W}^{K}$, and $\mathbf{E}_{\mathcal{S}}^{*}\mathbf{W}^{V}$, where $\mathbf{E}_{\mathcal{S}}^{*}$ is any sequence embedding defined in Eq.~(\ref{eq:sequence_embed}). The self-attention adopts the scaled dot-product attention~\cite{vaswani2017attention} on the query and key to measure weights from previous steps on the current step. 
% which can be formulated as:
% \begin{equation}
%     \text{Self-Attention}(\mathbf{Q}, \mathbf{K}, \mathbf{V}) = \sigma(\frac{\mathbf{Q}^T\mathbf{K}}{\sqrt{d}})\mathbf{V},
% \end{equation}
% where $\sigma(\cdot)$ denotes the softmax function.
%  However, the numerical scale of query, key, and value might fluctuate significantly. To avoid it, 
 To maintain numerical stability of query, key, and value, we apply ``exponential linear unit''~(elu) activation~\cite{clevert2015fast} after the linear transformations, which presents the distribution-based self-attention as:
\begin{equation}
    \textbf{DSA} = \sigma\left(\frac{\text{elu}(\mathbf{E}_{\mathcal{S}}^{*}\mathbf{W}^{Q})^T\text{elu}(\mathbf{E}_{\mathcal{S}}^{*}\mathbf{W}^{K})}{\sqrt{d}}\right)\text{elu}(\mathbf{E}_{\mathcal{S}}^{*}\mathbf{W}^{V}),
\end{equation}
where $\sigma(\cdot)$ denotes the softmax function and `DSA' is short for \textbf{D}istribution \textbf{S}elf-\textbf{A}ttention. We apply the DSA module in the mean Transformer as $\textbf{DSA}^{\mu}$ and the covariance Transformer as $\textbf{DSA}^{\Sigma}$, either taking $\mathbf{E}_{\mathcal{S}}^{*}=\mathbf{E}_{\mathcal{S}}^{\mu}$ or $\mathbf{E}_{\mathcal{S}}^{*}=\mathbf{E}_{\mathcal{S}}^{\Sigma}$ in Eq.~(\ref{eq:sequence_embed}) as inputs, respectively.  

\subsection{Mean and Covariance Feed-Forward Layers}
% Transformer introduces non-linearity and interactions on embedding dimensions with a point-wise feed-forward network. The network applies two layers of feed-forward networks on each item, with shared parameters. 
In additional to uncovering sequential patterns, we introduce two novel and adaptive versions of feed-forward layers 
% for better reprsentations at every position with
to endow extra non-linearity. 
We adapt the FFN layer to distribution representations by replacing the ReLU activation as ELU. 
% The proposed feed-forward layers adapt to distribution representations with the replacement of original ReLU with ELU activation function.
The FFN with respect to both $\textbf{DSA}^{\mu}_i$ and $\textbf{DSA}^{\Sigma}_i$ at position $i$ are defined as:
\begin{equation}
    \begin{aligned}
        \textbf{FFN}^{\mu}(\textbf{DSA}^{\mu}_i) &= \text{elu}\left(\text{elu}\left(\textbf{DSA}^{\mu}_iW^{\mu}_1+b^{\mu}_1\right)W^{\mu}_2+b^{\mu}_2\right),\\
        \textbf{FFN}^{\Sigma}(\textbf{DSA}^{\Sigma}_i) &= \text{elu}\left(\text{elu}\left(\textbf{DSA}^{\Sigma}_iW^{\Sigma}_1+b^{\Sigma}_1\right)W^{\Sigma}_2+b^{\Sigma}_2\right),
    \end{aligned}
\end{equation}
respectively,
where all $W$ are in $\mathbb{R}^{d\times d}$ and all $b$ are in $\mathbb{R}^d$. 

We retain those
% similar 
techniques used in existing Transformer structures, 
% for stabilizing the training process and avoiding over-fitting in both \textbf{DSA} and \textbf{FFN} layers
, including residual connection, dropout operation and layer normalization. Note that we can stack multiple \textbf{DSA} and \textbf{FFN} layers to learn more complex item relationships. In the meantime, the multi-heads mechanism is also applicable in \modelname framework.

\subsection{Layer Outputs}
However, a valid distribution requires covariance matrix to be positive definite while the outputs from {FFN} layers do not guarantee this property. Thus, we add an all ones vector to the output covariance embedding vector after the ELU activation, which is defined as follows:
\begin{equation}
    \Sigma_i = \mathrm{diag}\left(\text{elu}\left(\mathbf{O}^{(L)}_i\right) + \mathbf{1}\right),
\end{equation}
where $\mathbf{O}^{(L)}_i$ denotes the output covariance embedding of item $i$ after $L$ layers of $\textbf{DSA}^{\Sigma}$ and $\mathbf{FFN}^{\Sigma}$, and $\mathbf{1}\in \mathbb{R}^d$ is an all ones vector.

\subsection{Loss and Optimization}
Recall that the mean and covariance Transformers' outputs infer the mean and covariance embeddings of the next-step item at each position in the sequence, respectively. Following the same setting in~\cite{kang2018self}, we use the next-step item as ground truth prediction for each position in the sequence. For example, given a sequence $\mathcal{S}^{u}=[v^{u}_1, v^{u}_2, v^{u}_3, \dots, v^{u}_n]$, the ground truth item for $v^{u}_1$ is $v^{u}_2$. Note that if $v^{u}_i$ is a padding item, its ground truth item is also a padding item.

\subsubsection{Wasserstein Distance and Loss}
To measure how accurately the model predicts the next item, we need to identify the distance between the distribution of the ground truth item and the inferred distribution. The most popular approaches to calculate the distance between two distribution are Kullback–Leibler~(KL) divergence~\cite{kullback1951information} and $p_{th}$ Wasserstein distance~\cite{arjovsky2017wasserstein}. 
% However, KL divergence fails to fulfill the triangular inequality~\cite{kl_div_triangle} while \cite{verify_triangle} verifies that Wasserstein distance $W_p$ satisfies the triangle inequality. Moreover, 
Wasserstein distance can measure distances when two distributions have no overlap while KL divergence cannot. Therefore, we propose to use $p_{th}$ Wasserstein distance to measure the distance of Gaussian distributions. To be specific, we use 2-Wasserstein distance. Given two items $i_1$ and $i_2$, as well as their distrbutions $\mathcal{N}(\mu_{i_1}, \Sigma_{i_1})$ and $\mathcal{N}(\mu_{i_2}, \Sigma_{i_2})$, the Wasserstein distance is:
\begin{equation}
    d_{W_2}(i_1, i_2) = ||\mu_{i_1}-\mu_{i_2}||^2_2 + \text{trace}\left(\Sigma_{i_1}+\Sigma_{i_2}-2(\Sigma_{i_2}^{1/2}\Sigma_{i_1}\Sigma_{i_2}^{1/2})^{1/2}\right).
\end{equation}

\subsubsection{Loss}
Based on 2-Wasserstein distance $W_2$, we propose to use a Wasserstein distance-based BPR loss~\cite{rendle2012bpr} to measure the correctness of next-item prediction:
\begin{equation}
    -\sum_{\mathcal{S}^u\in\mathcal{S}}\sum_{t\in[1,2,\dots,|\mathcal{S}^u|]}\log (\sigma(d_{W_2}(i^{\prime}_t, \hat{t}) - d_{W_2}(i_t, \hat{t}))) + \lambda||\Theta||_2^2,
\end{equation}
where $\hat{t}$ denotes the inferred distribution at position $t$, $i_t$ is the ground truth item and $i^{\prime}_{t}$ denotes the negative sampled item from items that user $u$ never interacts with. $\Theta$ is the set of all learnable parameters in the both mean and covariance Transformers.

For evaluation, for user $u$, we calculate distance scores based on the 2-Wasserstein distance $W_2(\hat{\mathcal{S}^u_n}, i)$ on all candidate items ${i\in \mathcal{V}}$, where $\hat{\mathcal{S}^u_n}$ denotes the inferred next-item distribution at last position $n$. Then we rank them in ascending order to generate the recommendation list.

\section{Experiments}
In this section, we validate the effectiveness of the proposed \modelname by presenting experimental settings and results. The designed experiments will answer the following research questions:
\begin{itemize}[leftmargin=*]
    \item \textbf{RQ1}: Does \modelname outperform the state-of-the-art recommendation methods?
    \item \textbf{RQ2}: Does distribution representation provide better recommendations than single item embedding?
    \item \textbf{RQ3}: Are distribution representations in sequential modeling effective for alleviating cold-start user\slash item problems?
\end{itemize}

% Please add the following required packages to your document preamble:
% \usepackage{booktabs}
\begin{table} %[!htp]
\caption{Datasets Statistics}
\label{tab:data_stat}
\begin{tabular}{@{}l|llll@{}}
\toprule
Dataset & \#users & \#items & \#actions & density \\ \midrule
\textit{Amazon Toys} & 57,617 & 69,147 & 410,920 & 0.010\% \\
\textit{Amazon Beauty} & 52,204 & 57,289 & 394,908 & 0.013\% \\
\textit{Amazon Games} & 31,013 & 23,715 & 287,107 & 0.039\% \\ \bottomrule
\end{tabular}
\end{table}

\subsection{Datasets}
We evaluate the proposed \modelname on three public benchmark datasets from Amazon review datasets~\cite{mcauley2015image}. Amazon datasets are known for high sparsity and having several categories of rating reviews. Details of datasets statistics are presented in Table~\ref{tab:data_stat}. Following~\cite{kang2018self, sun2019bert4rec}, we treat the presence of ratings as positive implicit feedbacks. We use the timestamps of each rating to sort the interactions of each user to generate the sequence. The most recent interaction is used for test and the last second one is used for validation.

\subsection{Experimental Settings}
\textbf{Evaluation Protocol.} We evaluate all models with three standard metrics, \textit{Recall@N}, \textit{NDCG@N}, and \textit{Mean Reciprocal Rank~(MRR)}. Recall@N measures the accuracy of retrieving relevant items in the top-N recommendation. NDCG@N also considers the ranked positions of retrieved relevant items in the top-N list. MRR measures the ranking performance of the entire recommendation list. We set N to be 1 and 5 for evaluation. For each user, we randomly sample 1,000 items without interaction with the user as negative items considering ranking efficiency.\\
\textbf{Baselines.} We compare \modelname with several state-of-the-art recommendation methods in three relevant categories: static methods with point embedding vectors, static metric learning methods, and sequential methods. We use 
% the state-of-the-art BPRMF~\cite{rendle2012bpr} and
LightGCN~\cite{he2020lightgcn} as the strong baseline with point embedding vectors. We also compare with BPRMF~\cite{rendle2012bpr}, but we omit it in the performance table because of its low values. For metric learning methods, we compare the most recent work DDN~\cite{zheng2019deep}, and SML~\cite{li2020symmetric}. For sequential recommendation methods, we adopt the metric learning-based TransRec~\cite{he2017translation} and Transformer-based SASRec~\cite{kang2018self} as baselines.
Note that TransRec belongs to both the metric learning framework and sequential recommendation. \\
\textbf{Hyper-parameter Settings.} For all baselines, we search the embedding dimension in $\{32, 64, 128\}$. As the proposed model has both mean and covariance embeddings, we only search for $\{16, 32, 64\}$ for \modelname for a fair comparison. We search the $L$-2 regularization weight from $\{0.0, 0.001, 0.005, 0.01, 0.05, 0.1\}$. The number of layers $L$ is selected from $\{1,2\}$. For all baselines specific hyper-parameters, we do not list all of them because of the space limitations. We grid search all possible combinations for all models and report the test set performance based on the best validation MRR result.

% Please add the following required packages to your document preamble:
% \usepackage{multirow}
% \usepackage{graphicx}
% \usepackage[normalem]{ulem}
% \useunder{\uline}{\ul}{}
\begin{table}[]
\centering
\caption{Performance Comparison in Recall@1, Recall@5, NDCG@5,and MRR. The best and second-best results are boldfaced and underlined, respectively.}
\label{tab:perf_table}
\resizebox{0.48\textwidth}{!}{%
\begin{tabular}{cc|ccccccc}
\hline
\textbf{Dataset} & \textbf{Metric} & \multicolumn{1}{c}{\textbf{LGCN}} & \multicolumn{1}{c}{\textbf{TRec}} & \multicolumn{1}{c}{\textbf{DDN}} & \multicolumn{1}{c}{\textbf{SML}} & \multicolumn{1}{c}{\textbf{SASRec}} & \multicolumn{1}{c}{\textbf{\modelname}} & \multicolumn{1}{c}{\textbf{Imp.}} \\ \hline
\multirow{4}{*}{\textit{Toys}} & Recall@1 & 0.0603 & 0.0622 & 0.0657 & 0.0666 & {\ul 0.0669} & \textbf{0.0886} & +32.4\% \\
 & Recall@5 & 0.1494 & 0.1566 & 0.1423 & 0.1435 & {\ul 0.1661} & \textbf{0.1762} & +6.1\% \\
 & NDCG@5 & 0.1060 & 0.1085 & 0.0951 & 0.0933 & {\ul 0.1179} & \textbf{0.1341} & +14.5\% \\
 & MRR & 0.1084 & 0.1096 & 0.0954 & 0.0938 & {\ul 0.1187} & \textbf{0.1348} & +13.5\% \\ \hline
\multirow{4}{*}{\textit{Beauty}} & Recall@1 & 0.0592 & 0.0404 & 0.0464 & 0.0412 & {\ul 0.0667} & \textbf{0.0774} & +16.0\% \\
 & Recall@5 & 0.1510 & 0.1144 & 0.1232 & 0.1258 & {\ul 0.1541} & \textbf{0.1652} & +7.2\% \\
 & NDCG@5 & 0.1060 & 0.0781 & 0.0840 & 0.0788 & {\ul 0.1121} & \textbf{0.1230} & +9.7\% \\
 & MRR & 0.1081 & 0.0825 & 0.0880 & 0.0797 & {\ul 0.1124} & \textbf{0.1229} & +9.3\% \\ \hline
\multirow{4}{*}{\textit{Games}} & Recall@1 & 0.0825 & 0.0809 & 0.0722 & 0.0776 & {\ul 0.1111} & \textbf{0.1238} & +11.4\% \\
 & Recall@5 & 0.2313 & 0.2106 & 0.2218 & 0.2061 & {\ul 0.2870} & \textbf{0.3064} & +6.7\% \\
 & NDCG@5 & 0.1584 & 0.1474 & 0.1422 & 0.1437 & {\ul 0.2014} & \textbf{0.2183} & +8.3\% \\
 & MRR & 0.1610 & 0.1513 & 0.1423 & 0.1439 & {\ul 0.1984} & \textbf{0.2152} & +8.4\% \\ \hline
\end{tabular}%
}
\end{table}

\subsection{Overall Comparison~(RQ1 and RQ2)}
We report the overall performance comparison between the proposed \modelname and baselines in Table~\ref{tab:perf_table}. The followings are our observations:
\begin{itemize}[leftmargin=*]
    \item \textbf{RQ1:} \modelname significantly outperforms all baselines in all three datasets across all evaluation metrics. The average relative improvements over the second-best baseline is 19.9\% on Recall@1, 6.7\% in Recall@5, 10.8\% in NDCG@5, and 10.4\% in MRR. These improvements demonstrate the effectiveness of distribution-based representations with uncertainty in sequential recommendation.
    \item \textbf{RQ2:} Compared with the point embedding representation sequential model SASRec, the proposed \modelname still achieves a great improvement margin. The reasons for this improvement are twofold. First, the distribution-based representations provide uncertainty information when we model users with various interests. Moreover, the sequential modeling for mean and covariance can dynamically capture the evolving patterns of uncertainty within the sequence.
    \item Among all baselines, we can observe that sequential methods achieve outstanding performance, demonstrating the necessity of utilizing sequential information. Moreover, the state-of-the-art CF method LightGCN outperforms other static baselines, owing to its capability of using graph information.
\end{itemize}

\subsection{Performance w.r.t Sequence length~(RQ3)}
\begin{figure}[ht]
     \centering
     \begin{subfigure}[b]{0.235\textwidth}
         \centering
        \includegraphics[width=1\textwidth]{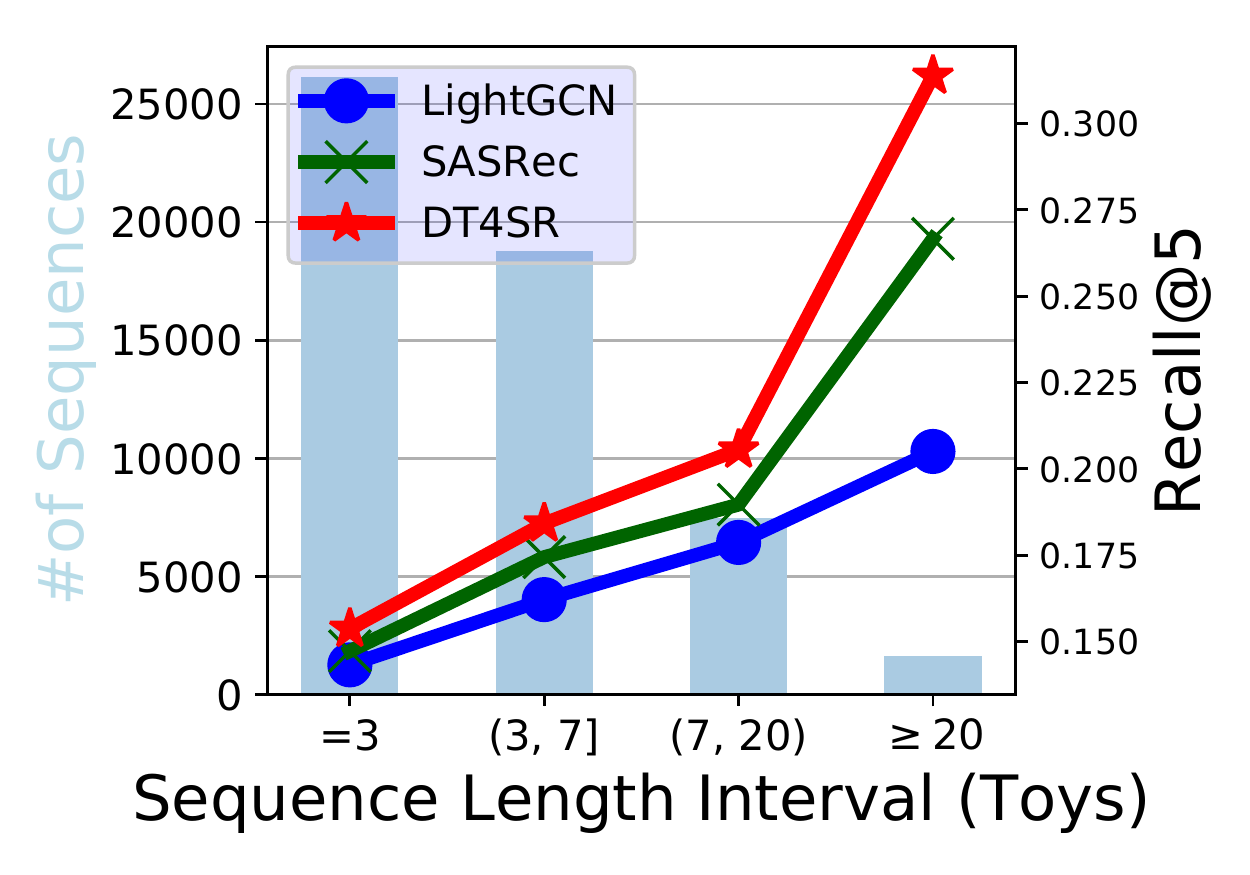}
        \label{fig:toys_all_recall_seqlength}
     \end{subfigure}
     \begin{subfigure}[b]{0.235\textwidth}
         \centering
        \includegraphics[width=1\textwidth]{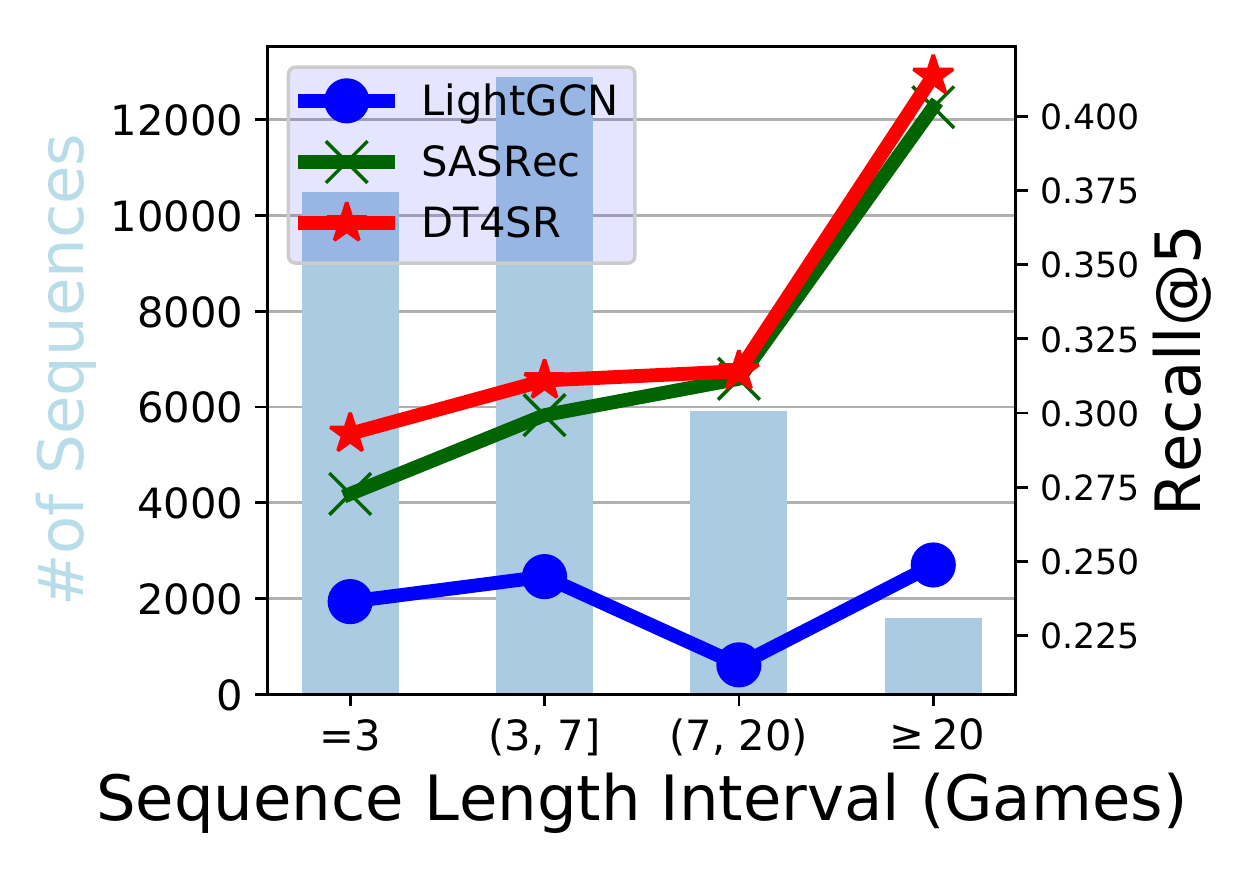}
        \label{fig:games_all_recall_seqlength}
     \end{subfigure}
     \vspace{-5mm}
     \caption{The Recall@5 performance on different sequence lengths over Amazon Toys and Amazon Games datasets.}
     \label{fig:perf_wrt_length}
\end{figure}
We present the performances of LightGCN, SASRec, and proposed \modelname with respect to different sequences lengths~(\textit{i.e.,} number of interactions of users) in Figure~\ref{fig:perf_wrt_length}. We can observe that \modelname performs the best on all sequence length intervals of users. Note that for users with only one interaction, the relative performance gain of \modelname against SASRec is 9\% on Recall@5. It demonstrates the effectiveness of \modelname in cold-start users. We can also observe that the long sequences~(\textit{i.e., }users with more than 20 interactions) achieve considerable improvements, especially on the Toys dataset. Long sequences typically cover diverse interests, and the observation proves the necessity of uncertainty in representing sequences. 

\subsection{Performance on Cold Start Items ~(RQ3)}
\begin{figure}[ht]
     \centering
     \begin{subfigure}[b]{0.235\textwidth}
         \centering
        \includegraphics[width=1\textwidth]{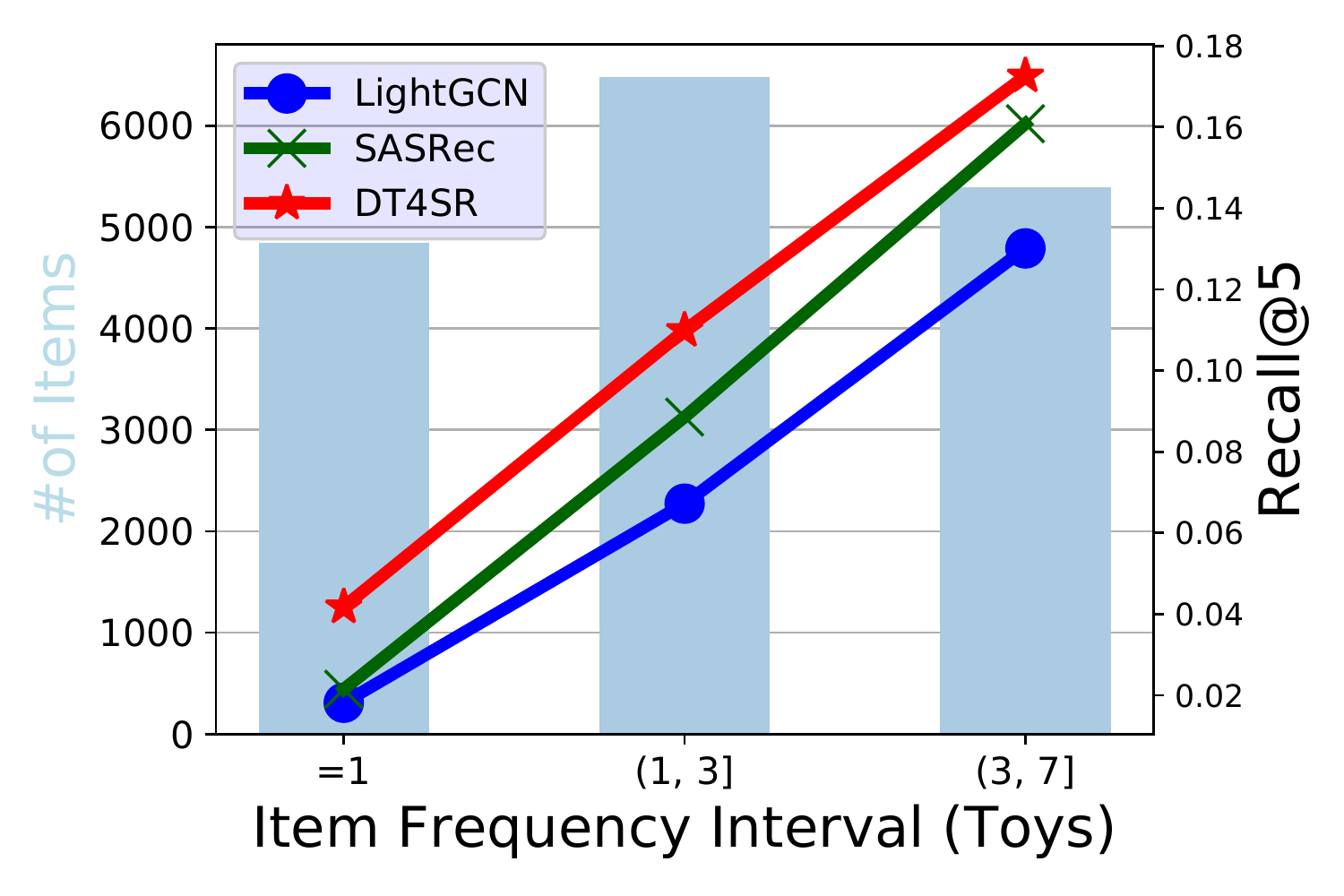}
        \label{fig:toys_all_recall_seqlength_colditem}
     \end{subfigure}
     \begin{subfigure}[b]{0.235\textwidth}
         \centering
        \includegraphics[width=1\textwidth]{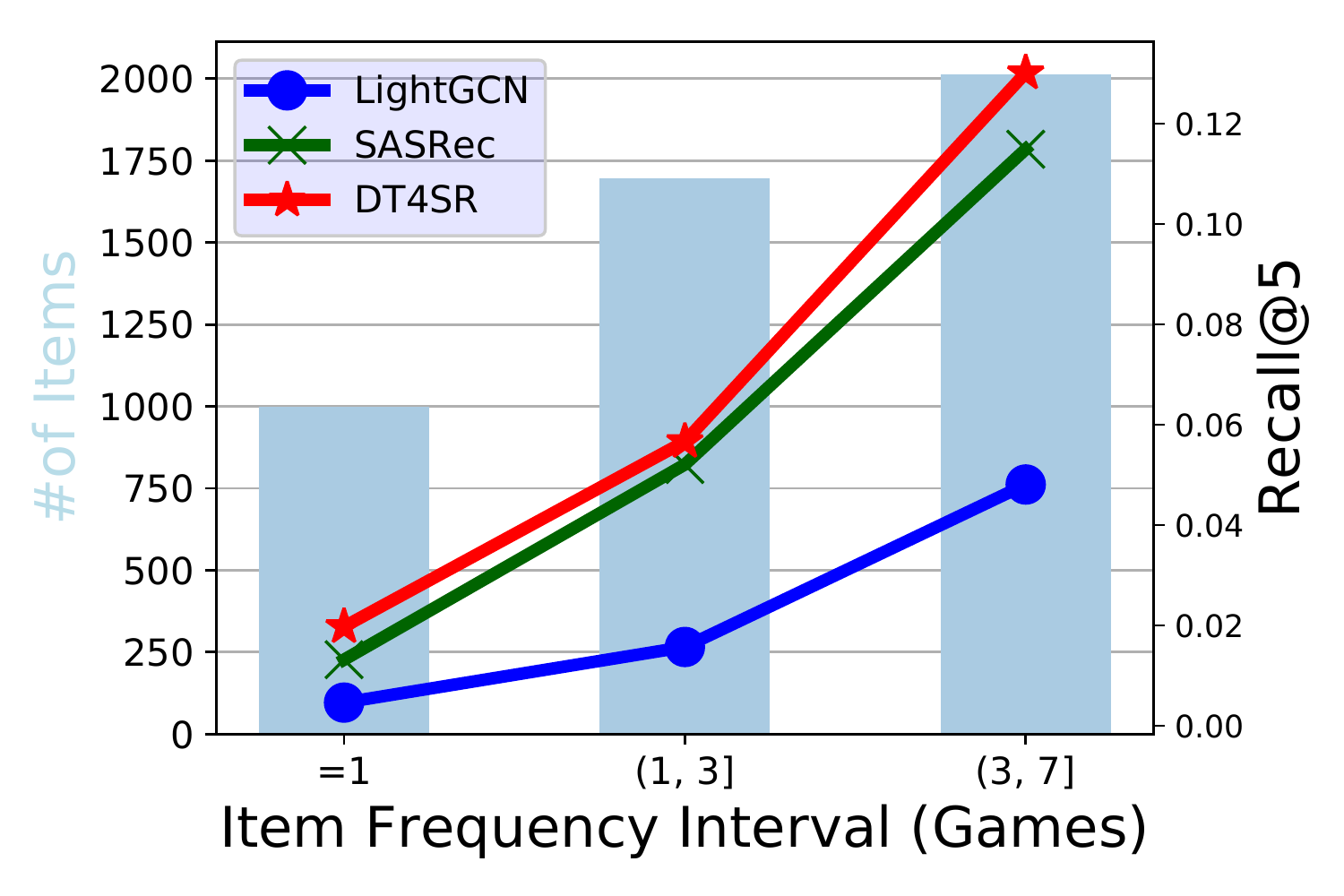}
        \label{fig:games_all_recall_seqlength_colditem}
     \end{subfigure}
     \vspace{-5mm}
     \caption{The Recall@5 performance on cold start items over Amazon Toys and Amazon Games datasets.}
     \label{fig:perf_wrt_colditems}
\end{figure}
We plot the performances of LightGCN, SASRec, and proposed \modelname on cold start items, which only have limited interactions in training data. For Toys dataset, \modelname achieves 100\% relative improvements on extremely cold start items~(\textit{i.e., } items with only one interaction). For the Games dataset, the performance gain is also more than 50\% for extremely cold start items. This experiment shows that distribution representations for items and sequences are more expressive and generalized for cold start items. Note that the \modelname still achieves comparative performance in frequent items.

\section{Conclusion}
We propose \modelname under the framework of Transformer with distribution representations in the sequential recommendation. Different from existing sequential methods assuming users are deterministic, we propose introducing uncertainty into representations of sequences and items. We use Elliptical Gaussian Distributions to represent items with uncertainty. Building upon distribution representations, we propose two novel Transformers for learning mean and covariance with the guarantee of positive definite property of covariance. We conducted several experiments to demonstrate that the proposed \modelname significantly outperforms existing methods in overall performance and cold start user\slash item recommendation.

%%
%% The acknowledgments section is defined using the "acks" environment
%% (and NOT an unnumbered section). This ensures the proper
%% identification of the section in the article metadata, and the
%% consistent spelling of the heading.

%%
%% The next two lines define the bibliography style to be used, and
%% the bibliography file.
\bibliographystyle{ACM-Reference-Format}
\bibliography{sample-sigconf}

%%
%% If your work has an appendix, this is the place to put it.
\end{document}